\documentstyle[aps,prd,preprint,tighten]{revtex}
\begin{document}
\draft
\title{Experimental Status of Quaternionic Quantum Mechanics}
\author{S.\ P.\ Brumby and G.\ C.\ Joshi.}
\address{School of Physics, University of Melbourne,
Research Centre for High Energy Physics,
\\Parkville, Victoria 3052, Australia.}
\maketitle
\begin{abstract}
Analysis of the logical foundations of quantum mechanics indicates the
possibility of constructing a theory using quaternionic Hilbert spaces.
Whether this mathematical structure reflects reality is a matter for 
experiment to decide.
We review the only direct search for quaternionic quantum mechanics yet
carried out and outline a recent proposal by the present authors to look
for quaternionic effects in correlated multi-particle systems.
We set out how such experiments might distinguish between the several 
quaternionic models proposed in the literature.
\end{abstract}
\draft
\pacs{{\it Report Number:} UM--P--94/120  ; RCHEP--94/34\\
{\it PACS:} 3.65Bz}
\narrowtext
\section{Quaternionic Theories.}
The established formalism of quantum mechanics identifies physical 
observables with Hermitian operators on a Hilbert space of states.
The existence of quantum states consisting of superpositions of pure
states is permitted by the Hilbert space structure.
Operators corresponding to different observables need not commute,
but the outcome of a measurement on a system will be an eigenvalue of
the corresponding Hermitian operator and hence real.
It is this last fact which led Birkhoff and von Neumann\cite{BvN} to
conclude that it is possible to consider the states to form a vector space 
over the real, complex or quaternionic algebras.
Birkhoff and von Neumann state that these theories cannot be differentiated
by known formalistic criteria.  

Stueckelberg \cite{Stk} has shown that the description of polarisation states
in a real number formulation of quantum mechanics (RQM) requires the
introduction of a super selection operator with all of the algebraic properties
of the imaginary unit $i$.
Hence, known phenomena have demanded the enlargement of RQM to be essentially
equivalent to complex quantum mechanics (CQM).

The remaining possibility involves considering Hilbert spaces over the
division algebra of real quaternions ${\bf H}$.
This is generated over the real numbers ${\bf R}$ by a basis of abstract 
elements $\{i_0 = 1,\,i_1 ,\,i_2 ,\,i_3 \}$ with the multiplication rule,
$\forall r \in \{1,\,2,\,3\}$,
\begin{equation}
1i_r = i_r 1 = i_r ,\ \ i_{1}^2 = i_{2}^2 = i_{3}^2 = -1\ \ \text{and}%
\ \ i_1 i_2 = -i_2 i_1 = i_3 .
\end{equation}
That is, any quaternion $a$ can be written in the form
\begin{equation}
a = a_0 + a_1 i_1 + a_2 i_2 + a_3 i_3 ,\ a_\mu \in {\bf R} .
\end{equation}
The algebra has the $\ast$-anti-involution $1^{\ast} = 1$, $i^{\ast}_r = -i_r$,
$(ab)^{\ast} = b^{\ast}a^{\ast}$, and a norm $\|a\|^2 = a^{\ast}a \in {\bf R}$.

If we introduce an additional complex unit $i_4 = \sqrt{-1}$, defined to
commute with each of $i_0$, $i_1$, $i_2$ and $i_3$, then the algebra ceases
to be a division algebra.  This complexification of ${\bf H}$ is called the 
biquaternionic algebra, and has been used to treat special relativity and
electromagnetism\cite{Lcz}.  The Pauli matrices are complex $2\times 2$ 
matrix representations of the Hermitian biquaternion units $i_4 i_r$.
The Lie algebra of ${\rm SU}(2)$ is ${\bf H}$ with the commutator product
$[a,b]=ab - ba$.  Hence, theories of spin-1/2 fermions,
Salam--Weinberg unified electro-weak theory, and Yang--Mills isospin can be
written in a quaternionic form\cite{IZ}.  For a review of these and other
uses of the quaternion algebra in standard quantum mechanics and quantum 
field theory, see \cite{Gsy} and references therein.

Quaternionic quantum mechanics (QQM) differs from these other
(mainstream) applications of the quaternions, because it considers the
Hilbert space of quantum states to be a vector space over ${\bf H}$.
Hence, eigenvalues of operators on the quaternionic Hilbert space
${\cal H}_{\bf H}$ need not commute, though we retain the identification  of
physical observables with Hermitian operators which, as is well known,
necessarily have real eigenvalues only.
The relation between kets of ${\cal H}_{\bf H}$ and quaternionic 
wavefunctions follows the usual pattern (where the $\Psi_\mu \in {\bf R}$).
\begin{equation}
\Psi(x_\mu ) \equiv \langle x_\mu |\Psi\rangle %
= \Psi_0 + \Psi_1 i_1 + \Psi_2 i_2 + \Psi_3 i_3 \,.
\end{equation}

Alternatively, we can express $\Psi$ as an ordered pair of $i_1$-complex
numbers, in its so-called symplectic representation,
\begin{equation}
\Psi(x_\mu ) = \psi_{\alpha} + i_2 \psi_{\beta} ,
\end{equation}
where $\psi_{\alpha} = \Psi_0 + i_1 \Psi_1$ and
$\psi_{\beta} = \Psi_2 - i_1 \Psi_3$.

If we define $|\vec{\Psi}| = (\sum_{r=1}^{3}\Psi_{r}^{2})^{-1/2} \in {\bf R}$,
then there exists a pure imaginary quaternion of norm unity $\eta$, such that
\begin{equation}
\Psi(x_\mu ) = \Psi_{0}(x_\mu ) + |\vec{\Psi}(x_\mu )|\eta(x_\mu ) .
\end{equation}
From this we see that any two wavefunctions of QQM do not commute unless
their imaginary parts are parallel.  This property has confounded the 
construction of a completely satisfactory tensor product, i.e., one which
would be quaternion linear in each factor, allow the definition of a tensor
product of operators on each factor, and admitting a positive scalar product
for the purpose of second quantisation of the theory
(see \cite{NJ,HzR} for comprehensive discussions).

The obvious and prodigious success of CQM in describing physical systems 
demands that QQM must reproduce its results wherever CQM has been successful.
This led Finkelstein {\it et al.} \cite{FJSS} to propose a geometric
generalisation of CQM, producing what is in effect an ``almost complex
quantum mechanics'' \cite{Nak}.  Thus QQM is supposed to be 
complex in any neighbourhood of an observation, modelled by replacing the 
imaginary unit of CQM with a field of pure imaginary unit quaternions 
$i_{\bf x}$ over spacetime, and the language of fibre-bundles is used 
throughout their work.
Over larger distances, topological properties of the spacetime manifold
are permitted to manifest themselves in quantum mechanical expectation values.
Finkelstein {\it et al.} identify an intermediary limit between full
geometric QQM and standard CQM, which they call the ``electromagnetic limit''.
In this limit the Q--covariant derivative of the imaginary unit vanishes
and the Q--curvature term in the Lagrangian looks like that of EM.  The
result of this situation is no extra quaternionic degrees of freedom
in the wavefunction, and the theory is equivalent to a (complex) Abelian
gauge field theory.

Adler\cite{Adler} and Horwitz\cite{Hz} have investigated the case where
it is possible, by use of a quaternionic gauge transformation, to
choose all of the energy eigenkets of a system to be in the same complex
subset of ${\cal H}_{\bf H}$.  Then QQM effects are dependent upon the 
existence of new, hyper-complex components of the fundamental forces.

The search for these hypothetical extra components motivated the 
experimental test of QQM undertaken by Kaiser, George and Werner \cite{KGW}
in 1984.

\section{Interferometry test of QQM.}
The experiment of Kaiser {\it et al.} is based on a proposal of Peres
\cite{Peres}, who suggested that if a pair of potential barriers were to
possess additional hyper-complex components, then a particle (in Peres's 
proposal, a neutron) traversing the pair will experience a shift in the
phase of its wavefunction which will depend upon the order in which the
barriers are traversed.

Note that even though relativistic formulations of QQM have been proposed
\cite{FJSS,Adler}, this experiment limits itself to the special case of
classical, external fields acting on nonrelativistic matter-waves, as this 
is the situation amenable to investigation by particle interferometry.
The problem of how to quantise quaternionic potentials remains open,
while quantisation of the $i_{\bf x}$ field requires the overcoming of
obstacles similar to those attending the construction of a quantised gravity.

Peres's original suggestion contained the idea that the one should use
potential barriers with a large absorptive cross-section, which 
in the standard framework of CQM are modelled by potentials with 
large, non-vanishing imaginary components.  The idea behind this was that 
the imaginary number appearing in the potential might depend upon the 
material, so potentials corresponding to different barriers would not
commute, while having a large imaginary component might make it more likely
that the hyper-complex contribution to each potential is significant.  
The total phase shift experienced by a neutron of wavelength $\lambda$,
traversing a plate of thickness $D$ and refractive index $n$ is
\begin{equation}
\phi = \frac{2\pi}{\lambda}(n-1)D ,
\end{equation}
where for the thermal neutrons used in the experiment, $n$ is related to the 
coherent nuclear scattering length $b$ and the atom density $N$ of the barrier
material, by
\begin{equation}
n = 1 - \frac{\lambda^2 Nb}{2\pi}.
\end{equation}
Kaiser {\it et al.} decided that to achieve very high sensitivity to 
differences in the total phase shift caused by traversal of the apparatus
in different directions, the thickness of the slabs should be such that
\begin{equation}
\phi = -\lambda NbD \approx 10\,000^{\circ} .
\end{equation}
This ruled out the possibility of using materials with very high absorption
cross-sections.  Instead, they chose aluminium and titanium for their
differing chemical and nuclear properties, aluminium having a positive real
scattering amplitude while titanium has a negative real scattering amplitude,
though both have small absorption cross sections.  The experiment, at the
10-MW University of Missouri Research Reactor, used thermal neutrons of
wavelength $\lambda = 1.268\,\text{\AA}$ in a Bonse--Hart three crystal 
Laue--Laue--Laue interferometer (see \cite{KW} for a review of the techniques
of neutron interferometry).  

Kaiser's {\it et al.} result was that in changing the order of traversal
of the Al and Ti barriers, no phase difference was observed to 1 part in
$30\,000$.

A theoretical analysis of the two potential barrier experiment was later
undertaken by Davies and McKellar \cite{DMK} using the quaternionic 
one dimensional Schr\"{o}dinger equation of Adler \cite{Adler}.
This uses the symplectic
decomposition of the wavefunction to rewrite the quaternionic problem into
one of a pair of complex fields coupled through the hyper-complex components
of the system Hamiltonian.  In time independent form, this is the pair of
equations
\begin{equation}
\left(-\frac{{\rm d}^2}{{\rm d}x^2} + V_{\alpha}\right)\psi_{\alpha}%
 - V^{\ast}_{\beta}\psi_{\beta} = E\psi_{\alpha}
\end{equation}
\begin{equation}
\left(\frac{{\rm d}^2}{{\rm d}x^2} - V_{\alpha}\right)\psi_{\beta}%
 - V_{\beta}\psi_{\alpha} = E\psi_{\beta}
\end{equation}
Davies and McKellar attempted to treat the problem exactly but found the 
resulting barrier transmission and reflection coefficients to be too 
unwieldy to analyse.  Instead, they turned to numerical methods and 
confirmed that the transmission coefficient for two quaternionic, square
potential barriers of different heights will change by a phase when the 
system is traversed in the opposite sense.  This then forces us to
ask how this conclusion can be consistent with the null result
(to one part in $30\,000$) of Kaiser {\it et al.}.

Now the possibility remains that QQM modifies only some of the fundamental
forces. 
The experiment of Kaiser {\em et al.} would appear to severely
constrain any quaternionic component of the strong nuclear force, though
this experiment does not rule out the possibility that there might be a 
quaternionic component to the underlying chromodynamics.  Though QQM was
at one time considered as a possible framework for QCD, Adler has 
proposed that QQM might provide a natural framework of preonic physics
\cite{Adler}.

Alternatively, it might be that the experiment of Kaiser {\em et al.}\ 
is too simplistic, as it only involves a single fundamental force. 
One can conceive of a situation in which we associate a different complex
algebra with each of the fundamental forces, so that within each class
of interactions complex quantum theory suffices to describe the experimental
phenomena.

To rule out this possibility, Klein\cite{Klein} has suggested that we should
repeat the experiment with combinations of different fundamental forces
\footnote{Since this article first appeared, this experiment has been
attempted.  We refer the reader to a description of the preliminary results
in \cite{ALL}.}.
The simplest extension is to introduce a sequence of metal barriers and 
magnetic fields which interact with the neutron intrinsic spin.
Additionally, there are known phenomena associated with the terrestrial 
gravitational field which are subject to investigation by neutron 
interferometric methods.
The fact that such effects have been successfully taken into account in
previous interferometric experiments means that we can already severely
constrain the contribution of a quaternionic addition to the gravitational 
force, or else deduce that the gravitational force reached in the classical
limit from a quaternionic theory of quantised gravity, involves observables and
states constrained to lie in a complex subset of ${\bf H}$ coincident with
that selected by the strong force.
  
An experiment of the type described in \cite{Klein}, if it were to produce a
null result, would leave only the weak interaction as a possible force with
a quaternionic component.  We should then search for new effects in
neutrino physics (already suggested by Peres\cite{Peres}).
The extreme weakness of these interactions, however, renders very unlikely
the possibility of carrying out interferometric experiments of the 
type previously considered.

A positive outcome from Klein's experiment would give us some knowledge of the
particular form of QQM preferred by Nature.  If there was observed a pure
phase shift introduced by swapping the order of potential barriers, this would
imply that we live in the Q--flat limit of QQM (or that Q--curvature is 
negligible at the scale of the experiment, and perhaps also in the vicinity of
the Earth). Any more complicated observational result, in particular the 
production of qualitatively different interferometric patterns, would imply
the active presence of a Q--curvature.

Before moving on to consider experiments on correlated few body systems, 
we mention that Peres \cite{Peres} in fact suggested three 
different types of experiments.  Besides the experimentally realised 
interferometry test described above, he derived a universal relation between 
the scattering cross-sections of three coherent sources, taken singly and
pairwise.  Numerical violation of this relation would indicate the failure
of the complex description.
Additionally, he suggested the investigation of K$_s$ meson regeneration
by different media taken singly and pairwise.
To the best of our knowledge, neither possibility has been taken up by
experimental researchers.

\section{Experiments on multi-particle correlated systems.}
Note that in the previous discussion, the presence of quaternionic effects
depends upon the existence of hyper-complex components in the interaction
potentials of a physical system. 
Theoretical analysis of this situation requires the adoption
of a particular form for the quaternionic dynamics, which in the
non-relativistic Q--flat limit might be characterised by the assumption of QQM
essentially reducing to CQM in potential free regions of space.
Recently, Brumby, Joshi, and Anderson \cite{BJA} have questioned the need to
demand this integrability of the Q--curvature,
instead proposing a different class of experiments.

To this end, consider the variant due to Greenberger {\it et al.} \cite{GHSZ}
of Bohm's {\em gedankenexperiment}\/ test of the Einstein--Podolsky--Rosen 
programme.  This envisages the preparation of an entangled four body state
\begin{eqnarray}
|\Psi _{\text{GHSZ}}\rangle & = & \frac{1}{\sqrt{2}}\left(%
|+\rangle_{1}\!\otimes|+\rangle_{2}\!\otimes|-\rangle_{3}\!\otimes%
|-\rangle_{4} \right. \nonumber \\
&& \ \ \ \ \ \:-\ %
\left. |-\rangle_{1}\!\otimes|-\rangle_{2}\!\otimes|+\rangle_{3}\!\otimes%
|+\rangle_{4}\right) ,
\end{eqnarray}
whose constituents are permitted to propagate to space-like separations.
We then carry out simultaneous measurements of a component of spin of each
particle (using, say, sets of Stern--Gerlach magnets) and consider the product
of these values.  The quantum mechanical expectation value is thus
\begin{equation}
E_{\text{GHSZ}}(\{{\bf n}_k\}) = \langle \Psi _{\text{GHSZ}}|%
\bigotimes_{j=1}^{4} {\bf n}_j\cdot \bbox{\sigma}_j%
|\Psi _{\text{GHSZ}}\rangle ,
\end{equation}
where ${\bf n}_j$ gives the orientation of the $j^{\text{th}}$
Stern--Gerlach apparatus, and $\sigma_j^s$ is the $s^{\text{th}}$
Pauli matrix at the position ${\bf x}_j$ of that apparatus
(and so is a $2\times 2$, $i_{{\bf x}_j}$-complex matrix).

Assuming only that in the nonrelativistic limit the intrinsic spin of a
particle does not effect its propagation in free space, 
Brumby {\it et al.} now suggest that this expectation value will be
sensitive to the Q--curvature throughout the 2d region whose boundary 
intersects the four observation points.  Hence, geometric QQM predicts
experimental results unexplainable within the structure of standard CQM.
Our major difficulty lies in
the adoption of a tensor product of single particle states which adequately
handles the noncommutation of quaternionic wavefunctions and also permits us to
return to an equivalent complex quantum mechanical system through some natural
limiting process (we use the tensor product of quaternionic Hilbert 
modules developed by Horwitz and Razon \cite{HzR}).  While this question 
remains open, investigation of alternative structures has suggested
composition of quaternionic Hilbert spaces could be viewed as a lattice
theoretic problem \cite{NJ}.

It is potentially significant that we show geometric QQM to agree with CQM
in the case of a two body correlated system, essentially due to the
dimensionality of the space within which we align our experimental apparatus,
and the natural requirement that we use only relative angles when constructing
the operator corresponding to the simultaneous observation of components of
intrinsic spin.  This special ``hiding'' effect in two body systems is lost
in ${\cal N} \geq 3$ body systems, and hints that the apparent lack of 
evidence for QQM may be due to the subtlety of the theory.

Experimental investigations of violations of the Bell inequalities have
almost always used photons rather than electrons (see \cite{CS,Asp}).  
Because photons possess a single quantum of angular momentum, the
operator corresponding to a polarisation filter has real components
and QQM has no opportunity to manifest itself.
This situation is changed by the use of circularly polarised photons, in which
case the quantum mechanical prescription for calculating the expectation
values necessarily uses complex numbers, giving the quaternionic entangled
state a probability distribution sensitive to a changing $i_{\bf x}$--field
(hence, not predicted by CQM).  
Our prediction of quaternionic terms in multiparticle correlation experiments
provides a further motivation to the reasons given by Greenberger {\em et al.}
to undertake such experiments.

With this type of experiment, we are probing the structure of 
geometric QQM's bundle of quaternionic fields over spacetime
(equivalent to determining the integrability of the almost complex structure). 
This raises the
possibility that topological defects in the spacetime manifold will have
a new way of effecting the predictions of quantum mechanics in their 
vicinity.  

Work is in progress on a quaternionic quantum field theory which will allow
us to begin to investigate the consequences of the inclusion of such objects
\cite{BJ}.

\acknowledgements
One of the authors, S.P.B., acknowledges the support of an Australian 
Postgraduate Research Award.  G.C.J.\ was supported by the Australian
Research Council and the University of Melbourne.

\end{document}